# Comparison Latent Semantic and WordNet Approach for Semantic Similarity Calculation

I Wayan Simri Wicaksana[1], Bambang Wahyudi[2]
iwayan@staff.gunadarma.ac.id , bwahyudi@staff.gunadarma.ac.id
1. Gunadarma University
Jl. Margonda Raya 100, Depok, Indonesia
2. Gunadarma University
Jl. Margonda Raya 100, Depok, Indonesia

## Abstract

*Information exchange among many sources in Internet is more autonomous, dynamic and free. The situation drive difference view of concepts among sources. For example, word 'bank' has meaning as economic institution for economy domain, but for ecology domain it will be defined as slope of river or lake.*

*In this aper, we will evaluate latent semantic and WordNet approach to calculate semantic similarity. The evaluation will be run for some concepts from different domain with reference by expert or human.*

*Result of the evaluation can provide a contribution for mapping of concept, query rewriting, interoperability, etc.*

## 1. Introduction

Internet era has positive contribution, however some new problems occurred as well. The main issue is heterogeneity of information level such as syntactic, structure and semantic.

Diversity of semantic is more and more complex cause of a lot of parties can be participated as information provider and user in information exchange. Every party has different view of a concept. For example, *transportation* concept for one source will define as transport infrastructure such as road, street. However, other source will define as business activity, such as delivery service, moving service. The problem is how to calculate the difference or similarity of a concept among sources.

There are some approach for semantic similarity calculation, for example manual, statistic, latent semantic and WordNet.

Semantic similarity is a study about semantic relationship. Relation is an approach to consider the relatedness of two concepts. Simple relation only consider IS-A (hupernymy/hyponymy) relation.

Heterogeneity is a problem for interoperability. One of diversity is semantic heterogeneity. To solve the semantic heterogeneity is an important effort to achieve the interoperability between diversity sources.

This paper will compare to method that can be used to choose a method for calculation semantic similarity in interoperability. Currently, I can not find previous works that evaluate between WordNet and latent semantic to calculate semantic similarity.

The paper will compare latent semantic approach to semantic similarity based on WordNet by using path length and information content. The paper is organized as follows: the first section is to address the background. The second section is to introduce WordNet and latent semantic, the comparison discuss in third section. Forth section is to summary as conclusion.

## 2. Semantic Similarity

Semantic similarity calculation is a process that need cooperation from some domains such as language, psychology, computer. The first step to calculate the semantic similarity is refer to similarity of terminology or often called as label. The terminology can be class, property or instance. According to Euzenat [2], terminological approach can be based on string and language. In this paper, language based is implemented by using lexicons (WordNet) and latent semantic.

### 2.1. WordNet

WordNet is a semantic network database for English language which developed by University Princenton. Some versions of WordNet.have been developed in many languages.

Basic part of WordNet is synset which a set of synonym of a concept. Synsets are related in some model, such as hypernymy (type of), meronymy (part of) and antonymy (opposite word).

Semantic similarity in WordNet can be divided into two methods, which called path length and information content method. The path length method is to calculate number of node or relation between node in taxonomy. Shorter distance between two concepts have higher similarity. The advantage of path length is not depended to corpus statical and word distribution. The weakness is for taxonomy which has uniform distance. Some example equations [6] of semantic similarity using path length are Leacock-Chodorow, Resnik, Wu-Palmer. In this paper, Wu-Palmer equation will be implemented for evaluation.

Information content of a node is -*log* of amount of probability (calculated based on corpus frequent) of all words which has synset. If *p(x)* is probability of a instance of x, so information content of x is -log p(x). In this paper Jiang-Conrath equation is utilized for evaluation.

## 2.2. Latent Semantic

Latent Semantic Analysis (LSA) [4] is a theory and method to extract and provide context representation of a word by using statistic computation of corpus of text. The basic idea is to aggregate of all word context. Good LSA can reflect human knowledge in some way.

The simple method of LSA process as follows:
- Text representation in matrix, row is unique word and column is related document. In this step, matrix *{X}* will be produced.
- Next step, LSA conduct singular value decomposition (SVD) to matrix *{X}*. SVD compose the matrix to product of three matrix. One component of matrix explains original of entity row as a vector of orthogonal value, other matrix describes original column and the third matrix as diagonal matrix consist of scale value to three matrix. The process of matrix decomposition as *{X} = {W} {S} {P}*.

## 3. Comparison

The purpose of the evaluation is to compare some approaches as follows:
- WordNet method by using Wu Palmer equation based on path length.
- WordNet method by using Jiang Conrath equation based on information content.
- Latent semantic based on corpus from General Reading up to 1$^{st}$ year collage.
- Latent semantic based on corpus from Encyclopedia.

The evaluation refers to above research then compare to expert view as Recall, Precession and F-measure.

### 3.1. Evaluation Setup

There are some preparation steps of evaluation as below:
- Search or develop a tool for similarity calculation based on section three. In this evaluation we utilize on-line tool for WordNet from http://marimba.d.umn.edu, and latent semantic from http://lsa.colorado.edu.
- Consider number of domains and concepts for evaluation. In this evaluation, three domains are used, including transportation [1], book publication [8] and business [6]. The domains are taken from papers which have evaluation result of similarity based on expert.

- Similarity calculation has procedure as follow (1) calculate combination of all concepts between two domains by using four type approaches in section three, (2) filter the result of calculations by using threshold value, the value of threshold from 0.7 to 1.0 with interval is 0.05. The result is presented as ζ, (3) create table calculation based on expert or called reference table (β), (4) compare result of calculation to reference as Δ, and (5) calculate value of Recall *(=Δ/β)*, Precession *(= Δ / ζ )*, and F-Measure *(=2/((1/Recall)+(1/Precession)))*.

### 3.2. Evaluation Result

Result of experiment are presented at figure 1, 2 and 3.

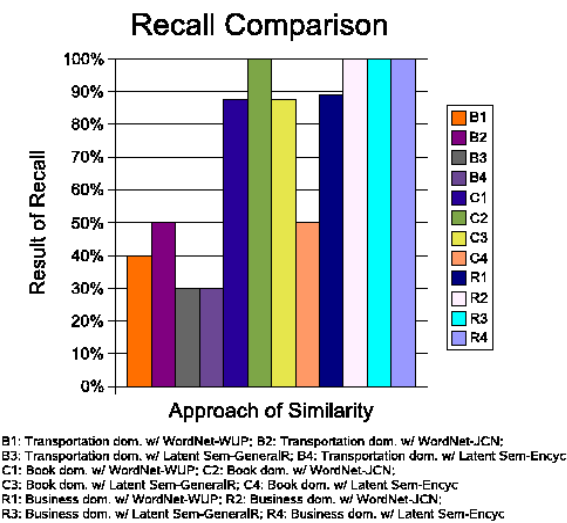

B1: Transportation dom. w/ WordNet-WUP; B2: Transportation dom. w/ WordNet-JCN;
B3: Transportation dom. w/ Latent Sem-GeneralR; B4: Transportation dom. w/ Latent Sem-Encyc
C1: Book dom. w/ WordNet-WUP; C2: Book dom. w/ WordNet-JCN;
C3: Book dom. w/ Latent Sem-GeneralR; C4: Book dom. w/ Latent Sem-Encyc
R1: Business dom. w/ WordNet-WUP; R2: Business dom. w/ WordNet-JCN;
R3: Business dom. w/ Latent Sem-GeneralR; R4: Business dom. w/ Latent Sem-Encyc

Figure 1. Recall Result

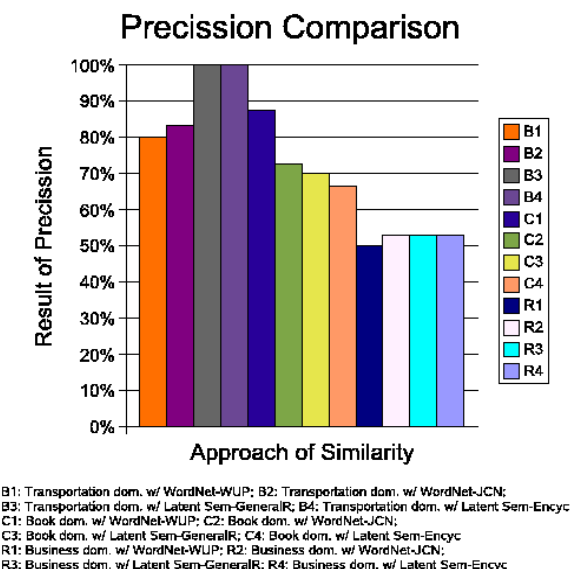

B1: Transportation dom. w/ WordNet-WUP; B2: Transportation dom. w/ WordNet-JCN;
B3: Transportation dom. w/ Latent Sem-GeneralR; B4: Transportation dom. w/ Latent Sem-Encyc
C1: Book dom. w/ WordNet-WUP; C2: Book dom. w/ WordNet-JCN;
C3: Book dom. w/ Latent Sem-GeneralR; C4: Book dom. w/ Latent Sem-Encyc
R1: Business dom. w/ WordNet-WUP; R2: Business dom. w/ WordNet-JCN;
R3: Business dom. w/ Latent Sem-GeneralR; R4: Business dom. w/ Latent Sem-Encyc

Figure 2. Precession Result

From three graphics, the result of evaluation as follows:

- At Recall, WordNet with Jean Conrath provide the best result at three domains, it is mean this approach able to calculate all similarities refer to expert, but there is a question is that all calculation is appropriate.

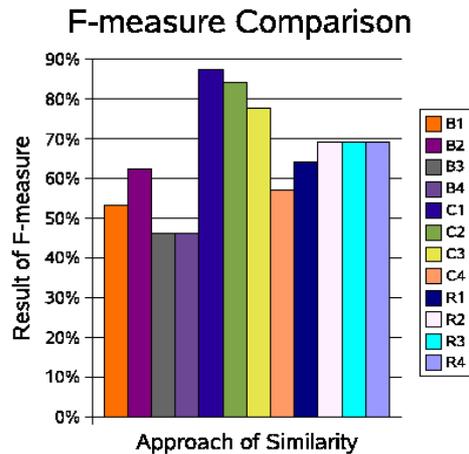

Figure 3. F-measure Result

- At Precession, there is no significant method can provide dominant result. Even though WordNet with Wu Palmer gave result little bit better than the others.
- At F-measure, WordNet with Wu Palmer has tendency better than the others.
- Refer to above result figure 3, the ranking of approach based on F-measure value (higher F-measure give better ranking position) refer to our evaluation are Wu-Palmer, Jean-Conrath, Latent Semantic (General Reading) and Latent Semantic (Encyclpedia). In some papers [5,7], Jean-Conrate equation is better than Wu-Palmer equation. The difference result can occurred caused of different version of WordNet and domain of evaluation.

## 4. Conclusion

The paper has provided evaluation of some approach for semantic similarity. The interesting result of evaluation is WordNet with path length is better than WordNet with information content and latent semantic. In summary, WordNet can provide better result of semantic similarity then latent semantic..

The main contribution of the paper is to provide information to select appropriate method for semantic web, ontology maintenance, mapping, and query rewriting. Choosing appropriate method is a difficult task, result of this experiment can be used as a references to decide the chosen method in semantic similarity measurement

In the future, the evaluation bring to more domains and concepts, the purpose to develop generic conclusion. Other thing, we plan to develop semi automatic tools to calculate semantic similarity from some domains.